%% file: main.tex
\documentclass[sigconf]{acmart}
\AtBeginDocument{%
  }

\setcopyright{acmlicensed}
\copyrightyear{2018}
\acmYear{2018}
\acmDOI{XXXXXXX.XXXXXXX}
\acmConference[Conference acronym 'XX]{Make sure to enter the correct
  conference title from your rights confirmation email}{June 03--05,
  2018}{Woodstock, NY}
\acmISBN{978-1-4503-XXXX-X/2018/06}

\usepackage{multirow}
\usepackage{arydshln}  
\usepackage{xspace}
\usepackage{pifont} 
\usepackage{graphicx}
\usepackage{tcolorbox}
\usepackage{subcaption}  
\usepackage{caption}  
\usepackage{hyperref}  
\definecolor{darkgreen}{rgb}{0.0, 0.5, 0.0}

\newcommand{\summaryb}[1]{%
\begin{tcolorbox}[
    colback=blue!5,  
    colframe=blue,  
    rounded corners, 
    arc=0.8mm,
    boxrule=0.2mm,  
    boxsep=0mm,    
    before skip=5pt,
    after skip=5pt,
]
#1
\end{tcolorbox}
}

\newcommand{\summaryc}[1]{%
\begin{tcolorbox}[
    colback=white!95!yellow,  
    colframe=yellow,  
    rounded corners, 
    arc=0.8mm,
    boxrule=0.2mm, 
    boxsep=0mm, 
    before skip=4pt,
    after skip=4pt,
]
#1
\end{tcolorbox}
}

\newcommand{\method}{AllianceCoder\xspace} 

\newcommand{\cmark}{\textcolor{darkgreen}{\ding{51}}} 
\newcommand{\xmark}{\textcolor{red}{\ding{55}}}   

\newcommand{\figmargin}{\vspace{-6pt}}
\newcommand{\tabmargin}{\vspace{-6pt}}

\begin{document}

\title{What Truly Matters? An Empirical Study on the Effectiveness of Retrieved Information in Retrieval-Augmented Code Generation}

\title{What to Retrieve for Effective Retrieval-Augmented Code Generation? An Empirical Study and Beyond}

\author{Wenchao GU}
\affiliation{%
  \institution{Technical University of Munich}
    \city{Heilbronn}
    \country{Germany}
}
\email{wenchao.gu@tum.de}

\author{Juntao Chen}
\affiliation{%
  \institution{Sun Yat-Sen University}
    \city{Zhuhai}
    \country{China}
}
\email{chenjt75@mail2.sysu.edu.cn}

\author{Yanlin Wang}
\affiliation{%
  \institution{Sun Yat-Sen University}
    \city{Zhuhai}
    \country{China}
  }
\email{yanlin-wang@outlook.com}

\author{Tianyue Jiang}
\affiliation{%
  \institution{Sun Yat-Sen University}
    \city{Zhuhai}
    \country{China}}
\email{jiangty9@mail2.sysu.edu.cn}

\author{Xingzhe Li}
\affiliation{%
  \institution{Sun Yat-Sen University}
    \city{Zhuhai}
    \country{China}}
\email{lixzh75@mail2.sysu.edu.cn}

\author{Mingwei Liu}
\affiliation{%
  \institution{Sun Yat-Sen University}
    \city{Zhuhai}
    \country{China}}
\email{liumw26@mail.sysu.edu.cn}

\author{Xilin Liu}
\affiliation{%
  \institution{Huawei Cloud Computing Technologies Co., Ltd}
    \city{Shenzhen}
    \country{China}}
\email{liuxilin3@huawei.com}

\author{Yuchi Ma}
\affiliation{%
  \institution{Huawei Cloud Computing Technologies Co., Ltd}
    \city{Shenzhen}
    \country{China}}
\email{mayuchi1@huawei.com}

\author{Zibin Zheng}
\affiliation{%
  \institution{Sun Yat-Sen University}
    \city{Zhuhai}
    \country{China}}
\email{zhzibin@mail.sysu.edu.cn}





\renewcommand{\shortauthors}{Gu et al.}


\input{sections/abstract}

\ccsdesc[500]{Software and its engineering~Automatic programming}
\ccsdesc[500]{Information systems~Information retrieval query processing}
\ccsdesc[500]{Computing methodologies~Natural language processing}

\keywords{Code Generation, LLM, Empirical Study}

\received{20 February 2007}
\received[revised]{12 March 2009}
\received[accepted]{5 June 2009}

\maketitle

\input{sections/introduction}
\input{sections/background}
\input{sections/experiment}
\input{sections/empirical}
\input{sections/methodology}
\input{sections/evaluation}
\input{sections/related_works}


\bibliographystyle{ACM-Reference-Format}
\bibliography{sample-base}










\end{document}

%% file: sections/abstract.tex
\begin{abstract}
Repository-level code generation remains challenging due to complex code dependencies and the limitations of large language models (LLMs) in processing long contexts. While retrieval-augmented generation (RAG) frameworks are widely adopted, the effectiveness of different retrieved information sources—contextual code, APIs, and similar snippets—has not been rigorously analyzed. Through an empirical study on two benchmarks, we demonstrate that in-context code and potential API information significantly enhance LLM performance, whereas retrieved similar code often introduces noise, degrading results by up to 15\%. Based on the preliminary results, we propose AllianceCoder, a novel context-integrated method that employs chain-of-thought prompting to decompose user queries into implementation steps and retrieves APIs via semantic description matching. Through extensive experiments on CoderEval and RepoExec, AllianceCoder achieves state-of-the-art performance, improving Pass@1 by up to 20\% over existing approaches. 
This study provides an experimental framework to further exploring \emph{what to retrieve} in RAG-based code generation, with our replication package available at \url{https://anonymous.4open.science/r/AllianceCoder} to facilitate future research.
\end{abstract}

%% file: sections/introduction.tex
\section{Introduction}


In modern software development, code generation~\cite{svyatkovskiy2020intellicode, zhou2023codebertscore, shin2021survey, li2022competition} has emerged as a critical capability to bridge the gap between natural language requirements from developers and executable code. With the advent of large language models (LLMs), state-of-the-art models have demonstrated impressive ability to generate standalone code snippets from natural language queries. However, these achievements are primarily limited to function-level code generation. Repository-level code generation - the task of generating code that with the repository context -remains a challenging task due to the complexity of code dependencies in repositories and inherent limitations of LLMs with long context. 
To address this challenge, most existing approaches~\cite{wu2024repoformer, deng2024r2c2, liao20243, zhang2023repocoder, wang2024rlcoder} adopt a retrieval-augmented generation (RAG) framework, which retrieves relevant information from the repository, appends it to the query, and then feeds it into LLMs for code generation.
Existing works adopts different information sources in the retrieval phase. 
(1) RepoFormer~\cite{wu2024repoformer}, RepoMinCoder~\cite{li2024repomincoder}, and $\rm R^2C^2$-Coder~\cite{deng2024r2c2} utilize contextual information from the current file to enhance code generation. By incorporating relevant local context and references to third-party libraries, these methods enable LLMs to infer and complete the target function more effectively. (2) RepoFuse~\cite{liang2024repofuse} extracts relevant APIs by analyzing import statements, while $\rm A^3$CodeGen~\cite{liao20243} first generates pseudo-code and then uses it to retrieve potential APIs for code generation. The retrieved API information helps LLMs correctly handle dependencies and ensures that the generated code adheres to API specifications. (3) RepoCoder~\cite{zhang2023repocoder} retrieves code snippets similar to the incomplete code using an context window, whereas RLCoder~\cite{wang2024rlcoder} applies reinforcement learning to optimize the retrieval process. Providing similar code snippets offers examples of recurring patterns, which aid LLMs in generating logically consistent implementations.

Although these information sources are widely used in RAG-based repository-level code generation, their impact on LLM performance has not been thoroughly studied. This raises a critical question: \textbf{In RAG-based repository-level code generation, what information sources truly matter?}  

To address this question, we conduct a preliminary study to analyze how different types of retrieved information contribute to improving LLM performance in repository-level code generation. Our experimental results reveal that \emph{contextual information} within the current file and \emph{potential API information} play a crucial role in enhancing LLM performance. Moreover, their combination further improves model performance in repository-level code generation.

Interestingly, contrary to its effectiveness in repository-level code completion, retrieving similar code snippets does not always improve LLM performance in repository-level code generation. In some cases, it can even have a negative impact. This occurs because there is no guarantee that a functionally similar code snippet exists within the repository, and retrieving dissimilar code snippets may mislead the LLMs, ultimately reducing the quality of the generated code.

Based on these findings, we conclude that retrieving contextual information from both the current file and relevant APIs is essential for enhancing LLM performance in repository-level code generation. While contextual information within the same file is naturally available and does not require additional retrieval techniques, identifying the appropriate APIs to invoke remains a challenge, as LLMs often lack prior knowledge of them.

To address this issue, we propose a simple yet effective approach, \method, for retrieving potential APIs. \method consists of three key stages: repository API processing, query processing, and context-integrated code generation.

In the repository API processing stage, we leverage LLMs to generate natural language descriptions for all APIs within the repository. These descriptions are then encoded into representation vectors for API retrieval. The motivation behind converting code into natural language is to bridge the semantic gap between programming languages and natural language, thereby improving retrieval accuracy.

In the query processing stage, we guide LLMs with examples to decompose the user query into multiple detailed implementation steps. For each step, the model generates descriptions of the potential APIs that may be invoked. The resulting API description sequence is then refined by LLMs and encoded into representation vectors using the same pre-trained model as in the repository API processing stage.

In the context-integrated code generation stage, we utilize these API representation vectors to retrieve the most relevant APIs from the repository based on cosine similarity. The retrieved APIs are then appended to the contextual information and user query before being fed into LLMs for repository-level code generation.

We evaluate our approach on two public datasets, CoderEval and RepoExec, and the results show that our two-step retrieval method outperforms state-of-the-art (SOTA) models by up to 20\% in terms of the $\rm Pass@1$ metric.

Our contributions can be summarized as follows:

\begin{itemize}
    \item We conduct an empirical study to investigate how different types of retrieved information contribute to LLM performance in repository-level code generation, showing that contextual information and potential APIs play a crucial role in improving performance.
    \item We propose a simple yet effective approach named \method that guides LLMs to decompose functionality into detailed implementation steps and generate descriptions of potential APIs for each step. These descriptions are then used to retrieve relevant APIs.
    \item We conduct extensive experiments to evaluate our approach. The results demonstrate that our simple two-step retrieval method significantly outperforms SOTA baselines.
\end{itemize}

%% file: sections/background.tex
\section{Background}
\subsection{Retrieval-Augmented Generation}

Retrieval-augmented generation (RAG) enhances generative models by integrating external knowledge~\cite{lewis2020retrieval}. Traditional models, while coherent, often suffer from factual inaccuracies and knowledge gaps due to their reliance on fixed memory~\cite{zhang2023repocoder, wang2024rlcoder}. RAG addresses these limitations by incorporating a retrieval mechanism that fetches relevant information from large-scale knowledge sources, such as databases or document corpora, enabling more accurate and contextually grounded outputs.

The RAG framework includes a retriever, which ranks relevant information based on the query, and a generator, which conditions its output on both the query and retrieved content, ensuring linguistic fluency while incorporating external knowledge. This adaptability allows the model to stay current without expensive retraining.

RAG has been successfully applied in domains requiring factual accuracy and knowledge grounding, including question answering~\cite{asai2023self, yu2024auto}, scientific text generation~\cite{dong2024advanced}, and automated code synthesis~\cite{rani2024augmenting}. By bridging retrieval-based and generative models, RAG advances the development of more reliable, context-aware AI systems.

The RAG process can be formulated as follows:
\begin{equation} 
A = Generate(q, Retrieval(q, \{c_1, \dots, c_n\})) 
\end{equation}
\noindent where $q$ represents the user query, $\{c_1, \dots, c_n\}$ denotes the candidate knowledge set, and  $A$ is the output generated by the language model. The function $Retrieval(\cdot)$ retrieves relevant candidates from the knowledge base based on the given query, while $Generate(\cdot)$ combines the retrieved information with the original query and feeds it into the language model for response generation.

Repository-level code generation is particularly well-suited for the RAG framework. Software repositories contain extensive historical code, library functions, API calls, and contextual information, making it impractical to rely solely on user queries for code generation. Moreover, due to the maximum input length constraints of LLMs, it is infeasible to import an entire repository into the model. RAG overcomes these limitations by retrieving relevant code snippets from the repository, enriching the model’s context and improving the accuracy and coherence of generated code. Additionally, large codebases often contain implementations of similar functions. By retrieving and reusing high-quality existing code snippets, RAG ensures consistency in coding style and interface design while reducing redundancy and potential errors. Consequently, most LLM-based approaches for repository-level code generation incorporate RAG to enhance performance.

\subsection{Repository-level Code Generation}
Before the advent of LLMs, most research efforts in code generation were limited to standalone code snippets due to the constrained performance and short input length of deep learning models. However, with the rapid advancement of LLMs, both their code generation capabilities and input length have significantly improved. This progress has shifted the research focus toward repository-level code generation, which better aligns with real-world software development scenarios.

In this context, RAG frameworks have been widely adopted in LLM-based repository-level code generation approaches. The repository-level code generation process can be formulated as follows:
\begin{equation} Code = Generate(q, Retrieval(q, Codebase)) \end{equation}
\noindent where $Codebase$ represents the given repository, and $q$ is the user query describing the target function. The function $Retrieval(\cdot)$ extracts relevant information from the $Codebase$ based on the query, while $Generate(\cdot)$ integrates the retrieved information with the original query and utilizes LLMs to generate the desired code.

\begin{table}[t]
    \setlength\tabcolsep{8pt}
    \centering
    \small
    \caption{Types of retrieved information by different LLM-based approaches.} \label{tab:baselines}
    \tabmargin
    \begin{tabular}{lccc}
    \toprule
    \textbf{Approach} & \textbf{Similar Code} & \textbf{API} & \textbf{Context}\\
    \midrule
        $\rm A^{3}CodeGen$ &  \xmark & \cmark & \cmark \\
        $\rm RepoCoder$ &  \cmark & \xmark & \cmark \\
        $\rm RepoFormer$ &  \cmark & \xmark & \cmark \\
        $\rm RepoMinCoder$ &  \cmark & \xmark & \cmark \\
        $\rm RLcoder$ &  \cmark & \xmark & \cmark \\
        $\rm R^2C^2-Coder$ &  \cmark & \xmark & \cmark \\
        $\rm GraphCoder$ &  \cmark & \xmark & \cmark \\
        $\rm RepoFuse$ &  \cmark & \cmark & \cmark \\
    \bottomrule
    \end{tabular}
\end{table}

Table~\ref{tab:baselines} summarizes the types of retrieved information used in current LLM-based repository-level code generation approaches. These types can be broadly classified into three categories: similar code, API, and context.

\textbf{Similar code} refers to the similar code snippets retrieved from the same repository. The underlying assumption is that similar code shares certain characteristics with the target code, providing LLMs with useful hints for implementation. From Table~\ref{tab:baselines}, we observe that most LLM-based approaches retrieve similar code from the repository, with the exception of $\rm A^3CodeGen$.

\textbf{API} retrieval focuses on identifying relevant APIs that should be invoked in the target code. The goal is to reduce code redundancy while enabling customization and supporting complex functionality. Information about the available APIs helps capture partial functionality within the overall target implementation, improving code integration, reusability, and reducing implementation difficulty. Unlike similar code, which is widely used in existing approaches, only $\rm A^3CodeGen$, and $\rm RepoFuse$ retrieve relevant APIs from the repository.

\textbf{Context} refers to contextual information from the same file as the user query. It includes details such as library imports, class definitions, and member function implementations. This information helps LLMs better understand the code style, maintain consistency, reduce redundancy, and correctly interpret the target functionality. Since contextual information is inherently available within the same file as the user query (which typically corresponds to the target functionality description), all prior LLM-based approaches utilize this information for repository-level code generation.

\paragraph{Limitations.} Although current LLM-based code generation approaches widely retrieve such information, the impact of each type on model performance and the optimal way to integrate them remain unclear.

%% file: sections/experiment.tex
\section{Experimental Setup}
In this section, we introduce the datasets, baselines and LLMs being evaluated in the experiments, and the evaluation metrics.

\subsection{Dataset}
Our experiments are conducted on \textbf{CoderEval}~\cite{yu2024codereval} and \textbf{RepoExec}~\cite{hai2024impacts}. CoderEval is a benchmark used to evaluate code generation performance on pragmatic code generation tasks. It consists of 230 Python and 230 Java tasks from real-world open-source projects. Each task contains a function signature, a function description, a solution, and several unit tests to assess the functional correctness of the generated code. RepoExec is a benchmark designed to evaluate repository-level code completion with complex contextual dependencies. It assesses models' ability to generate executable and functionally correct code while utilizing cross-file contexts. Each task provides essential code dependencies specified by developers, along with comprehensive test cases to verify functional correctness.

\subsection{Baselines and LLMs evaluated}
We explore two state-of-the-art frameworks for repository-level code generation: \textbf{RepoCoder}~\cite{zhang2023repocoder} and \textbf{RLCoder}~\cite{wang2024rlcoder}. RepoCoder is an iterative retrieval-generation pipeline. It effectively utilizes information scattered across different files within a repository and can generate code at various levels of granularity. RLCoder enhances this approach by implementing a novel reinforcement learning framework that enables the retriever to learn useful content without labeled data. It evaluates retrieved content based on perplexity metrics and includes a stop signal mechanism to determine when to retrieve and which candidates to retain. 
Both frameworks demonstrate significant improvements over baseline methods in repository-level code generation tasks.

\subsection{Embedding Model}
UniXcoder~\cite{guo2022unixcoder} is a dense retriever, encoding both queries and code snippets into dense vector representations. This vectorization mechanism enables the efficient discovery and extraction of semantically aligned code fragments from extensive repositories, based on the computational similarity between their respective vector embeddings.

\subsection{Metrics}
Since multiple implementations can achieve the same functionality, we adopt $Pass@k$~\cite{chen2021evaluating} as our sole evaluation metric. In our evaluation, a generated code snippet is considered successful only if it passes all test cases. The $Pass@k$ metric represents the success rate after generating code $k$ times using LLMs.

\subsection{Implementation Details}
In all our experiments, we utilized GPT-4o Mini and Gemini 1.5 Flash with the default temperature setting of 0.7. In the preliminary study on similar code retrieval, we selected the top five most similar code snippets for code generation. For our proposed method, \method, we retrieved only the single API from the repository that exhibited the highest cosine similarity to each API description generated by the LLMs.

%% file: sections/empirical.tex
\section{Preiliminary Study}

To investigate the impact of different information types on the LLM's ability to generate repository-level code under the RAG framework, we conduct an preiliminary study. In this study, we examine three types of information: contextual information, relevant code (i.e., code similar to the target code), and invoked APIs. To assess the upper-bound performance of each information type, we assume access to perfect information.

\begin{itemize}
\item \textbf{Contextual information}: This information type includes all text preceding the target function within the same file, such as library import statements, class definitions, and other implemented functions. By providing rich contextual information, it helps the LLM better understand the target function's environment. The experimental results for this setting are denoted as $\rm Context_{LLM}$.
\item \textbf{Relevant code}: Building on previous model-based approaches, we leverage LLMs to encode all code candidates within the repository into vector representations. Assuming prior knowledge of the correct target code, we adopt the retrieval framework used in RepoCoder~\cite{zhang2023repocoder}, which utilizes a context window to retrieve the top five most relevant code snippets based on vector similarity. The experimental results for this setting are denoted as $\rm Similar_{LLM}$.
\item \textbf{Invoked APIs}: We assume prior knowledge of all APIs invoked within the target code and retrieve them to facilitate repository-level code generation. The experimental results for this setting are denoted as $\rm API_{LLM}$.
\end{itemize}

In our experiments, we evaluate all possible combinations of the three types of retrieved information. Specifically, $\rm ConSim_{LLM}$ represents the performance of the LLM when provided with both contextual information and relevant code as input, while $\rm ConAPI_{LLM}$ corresponds to the setting where the LLM receives contextual information and the invoked API. Similarly, $\rm SimAPI_{LLM}$ denotes the case where the LLM is given relevant code and the invoked API. Finally, $\rm ConSimAPI_{LLM}$ represents the setting in which the LLM is provided with all three types of information: contextual information, relevant code, and the invoked API.

To assess the inherent capability of LLMs in repository-level code generation, we also conduct an experiment in which only the user’s original query is fed into the model without any additional information. The results of this setting are denoted as $\rm Pure_{LLM}$.

Overall, our preliminary study evaluates eight different settings, each corresponding to a unique combination of retrieved information types.

\section{Results of Preliminary Study}

In this section, we aim to answer the following research questions(RQs):

\begin{itemize}
    \item \textbf{RQ1: How does each type of information contribute to performance improvement in repository-level code generation?}
    \item \textbf{RQ2: Can the performance gains from contextual information and invoked APIs be fully captured by simply integrating them?}
    \item \textbf{RQ3: Are the performance improvements from contextual information and invoked APIs independent of each other?}
\end{itemize}

\begin{table}[t]
    \setlength\tabcolsep{1.5pt}
    \centering
    \small
    \caption{Preliminary experimental results on LLM repository-level code generation performance with different retrieved information types (best performance in bold).}
    \label{tab:preiliminary}
    \tabmargin
    \begin{tabular}{lrrrrrr}
    \toprule
    \multirow{2}{*}{\textbf{Method}} & \multicolumn{3}{c}{\textbf{RepoExec}} & \multicolumn{3}{c}{\textbf{CoderEval}}\\
    \cmidrule(r){2-4} 
    \cmidrule(l){5-7}
    & \textbf{Pass@1} & \textbf{Pass@3} & \textbf{Pass@5} & \textbf{Pass@1} & \textbf{Pass@3} & \textbf{Pass@5} \\
    \midrule
        $\rm Pure_{GPT}$ & 16.62 & 19.72 & 23.66 & 18.26 & 25.22 & 26.52\\
        $\rm Context_{GPT}$ & 29.01 & 34.08 & 38.59 & 30.43 & 36.52 & 37.83 \\
        $\rm Similar_{GPT}$ & 16.90 & 23.38 & 24.79  & 20.00 & 23.04 & 23.48 \\
        $\rm API_{GPT}$ & 25.63 & 32.39 & 34.93 & 25.65 & 30.43 & 33.04 \\
        $\rm ConSim_{GPT}$ & 22.53 & 26.48 & 27.60 & 14.35 & 20.43 & 23.04 \\
        $\rm ConAPI_{GPT}$ & \textbf{37.75} & \textbf{44.51} & \textbf{47.04} & \textbf{36.52} & \textbf{41.30} & \textbf{41.47}\\
        $\rm SimAPI_{GPT}$ & 25.36 & 31.55 & 34.09 & 22.17 & 26.52 & 26.52\\
        $\rm ConSimAPI_{GPT}$ & 29.86 & 38.31 & 38.87 & 19.57 & 23.91 & 26.52\\
    \midrule
        $\rm Pure_{Gemini}$ & 16.90 & 20.28 & 23.10 & 17.39 & 21.74 & 25.22\\
        $\rm Context_{Gemini}$ & 30.99 & 34.08 & 35.77 &23.04 & 26.09 & 27.39 \\
        $\rm Similar_{Gemini}$ & 17.18 & 22.25 & 23.38 & 13.48 & 16.52 & 16.96 \\
        $\rm API_{Gemini}$ &  26.20 & 30.99 & 34.93 & 22.61 & 24.78 & 26.09 \\
        $\rm ConSim_{Gemini}$ &  24.22 & 27.33 & 29.02 & 13.91 & 15.22 & 15.22\\
        $\rm ConAPI_{Gemini}$ & \textbf{38.59} & \textbf{41.69} & \textbf{44.51} & \textbf{25.22} & \textbf{27.39} & \textbf{28.70}\\
        $\rm SimAPI_{Gemini}$ & 27.32 & 31.27 & 32.96 & 19.13 & 22.17 & 22.61 \\
        $\rm ConSimAPI_{Gemini}$ & 32.67 & 36.90 & 38.59 & 22.17 & 23.04 & 23.91\\
    \bottomrule
    \end{tabular}
\end{table}

\subsection{RQ1: How does each type of information contribute to performance improvement in repository-level code generation?}

Table~\ref{tab:preiliminary} presents the experimental results of our preiliminary study. As shown in the table, contextual informaion provides the most significant performance improvement for LLMs in repository-level code generation. This result is expected, as contextual information supplies critical supplementary information, including imported libraries, class definitions, and implementations of other member functions. These elements enrich the user query with additional context, enabling LLMs to better understand the intended functionality and generate more accurate code.

Moreover, we observe that incorporating invoked API also substantially enhances performance. This improvement stems from two key factors. First, explicitly specifying relevant APIs within the repository allows LLMs to invoke them directly rather than reimplementing functionality from scratch, thereby reducing the likelihood of incorrect implementations. Second, API chains provide a structured reference, guiding LLMs to follow an established logical flow, which simplifies the code generation process.

Furthermore, the combination of both contextual information and invoked API information yields the best overall performance. From a human developer’s perspective, these elements collectively provide all the essential knowledge required for repository-level code generation, enabling more efficient implementation of the target functionality. Similarly, LLMs can leverage this prior knowledge to enhance their generative capabilities. Given that modern LLMs already excel at function-level code generation, supplying contextual information and API usage information effectively decomposes the complex task of repository-level code generation into smaller, well-defined function-level subtasks. This decomposition allows LLMs to fully utilize their strengths in function-level generation, leading to more accurate and reliable repository-level code generation.

However, an unexpected finding is that the retrieved relevant code contributes very little to performance improvement and can even degrade performance when combined with other types of supplementary information. For instance, in both datasets, the model incorporating all three types of information (contextual information, invoked APIs, and relevant code) performs significantly worse than the model that utilizes only contextual information and invoked APIs. Notably, in the CodeEval dataset, the model leveraging all three types of information underperforms even when compared to models using only contextual information or only invoked APIs. In fact, its performance is close to that of LLMs without any retrieved information at all.

This degradation is likely due to the discrepancy between the retrieved similar code and the target code. Ideally, the retrieved relevant code should exhibit functional similarities with the target implementation, serving as a reference to guide LLMs in generating correct code. However, there is no guarantee that such suitable examples exist within the repository. In repository-level code completion tasks, where only a few lines need to be completed, the likelihood of finding similar patterns within the repository is relatively high. However, in repository-level code generation tasks, which require generating entire functions, it becomes significantly harder to retrieve functionally similar code from the repository. In practice, retrieved code snippets often fail to meet this requirement, providing misleading information instead. When LLMs rely on irrelevant or functionally dissimilar code, they are more likely to produce incorrect outputs, ultimately degrading performance rather than enhancing it.

\summaryc{\textbf{Finding 1:} Contextual information and invoked APIs contribute significantly to performance improvements. However, the inclusion of similar code does not always lead to improvements and, in some cases, can even negatively impact performance in repository-level code generation.}

\begin{figure*}[h!]
\centering
\begin{subfigure}{0.24\textwidth}
    \centering
    \includegraphics[width=\textwidth]{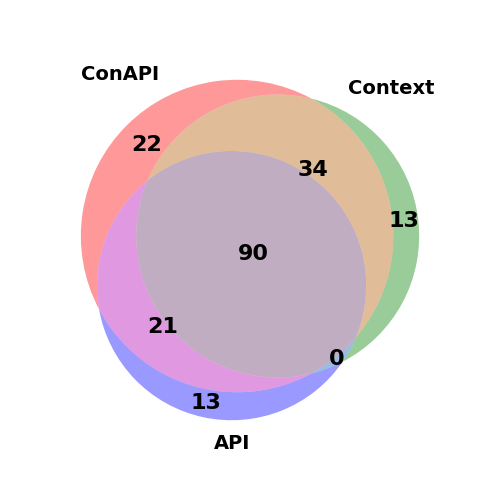}  
    \figmargin
    \captionsetup{width=0.95\textwidth}  
    \caption{Intersection of correct answers in RepoExec with GPT}
    \label{fig:intersection:repo_gpt}
\end{subfigure}%
\begin{subfigure}{0.24\textwidth}
    \centering
    \includegraphics[width=\textwidth]{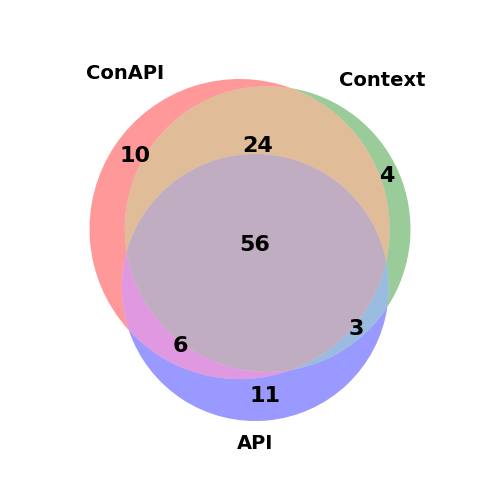}  
    \figmargin
    \captionsetup{width=0.95\textwidth}  
    \caption{Intersection of correct answers in CodeEval with GPT}
    \label{fig:intersection:code_gpt}
\end{subfigure}%
\begin{subfigure}{0.24\textwidth}
    \centering
    \includegraphics[width=\textwidth]{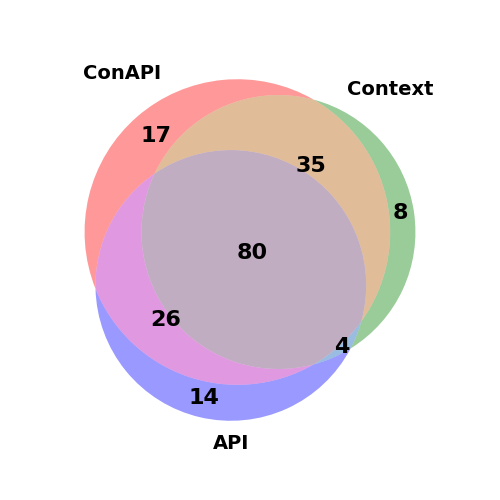}  
    \figmargin
    \captionsetup{width=0.95\textwidth}  
    \caption{Intersection of correct answers in RepoExec with Gemini}
    \label{fig:intersection:repo_gemini}
\end{subfigure}%
\begin{subfigure}{0.24\textwidth}
    \centering
    \includegraphics[width=\textwidth]{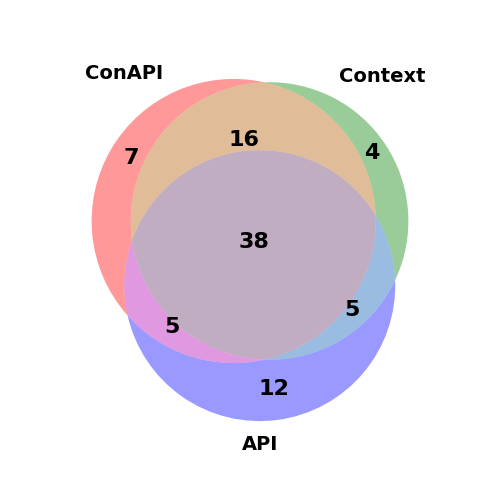}  
    \figmargin
    \captionsetup{width=0.95\textwidth}  
    \caption{Intersection of correct answers in CodeEval with Gemini}
    \label{fig:intersection:code_gemini}
\end{subfigure}
\figmargin
\caption{Intersection of correct answers across ConAPI, API, and Context under various LLMs and datasets.}
\label{fig:intersection}
\end{figure*}

\begin{figure*}[h!]
\centering
\begin{subfigure}{0.24\textwidth}
    \centering
    \includegraphics[width=\textwidth]{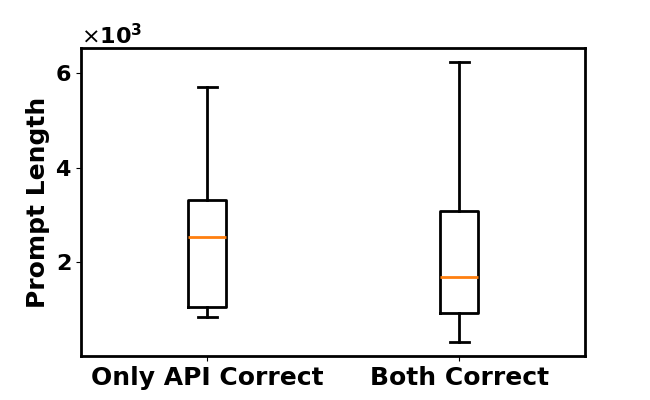}  
    \captionsetup{width=0.95\textwidth}  
    \caption{Comparison of performance based on input prompt length in RepoExec with GPT}
\end{subfigure}%
\begin{subfigure}{0.24\textwidth}
    \centering
    \includegraphics[width=\textwidth]{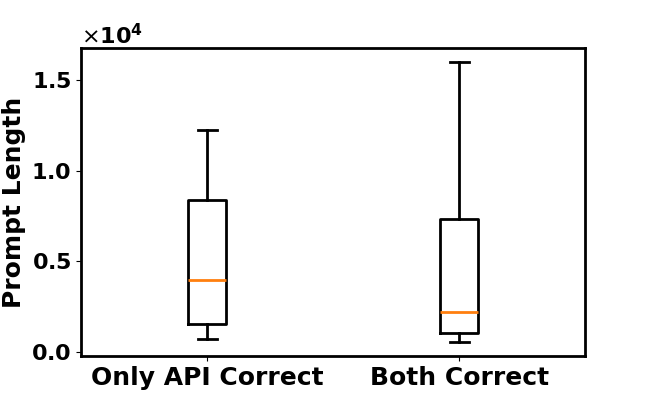}  
    \caption{Comparison of performance based on input prompt length in CodeEval with GPT}
\end{subfigure}%
\begin{subfigure}{0.24\textwidth}
    \centering
    \includegraphics[width=\textwidth]{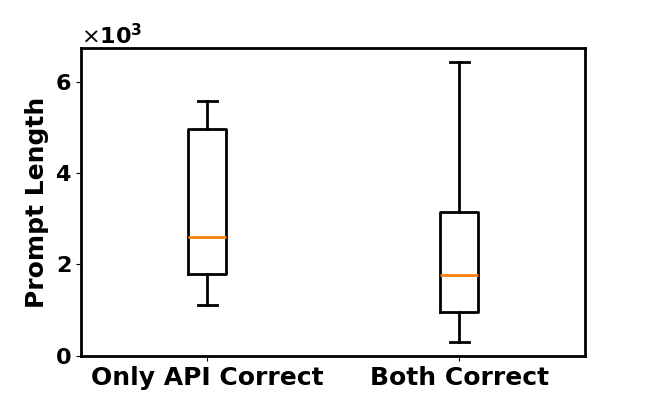}  
    \captionsetup{width=0.95\textwidth}  
    \caption{Comparison of performance based on input prompt length in RepoExec with Gemini}
\end{subfigure}%
\begin{subfigure}{0.24\textwidth}
    \centering
    \includegraphics[width=\textwidth]{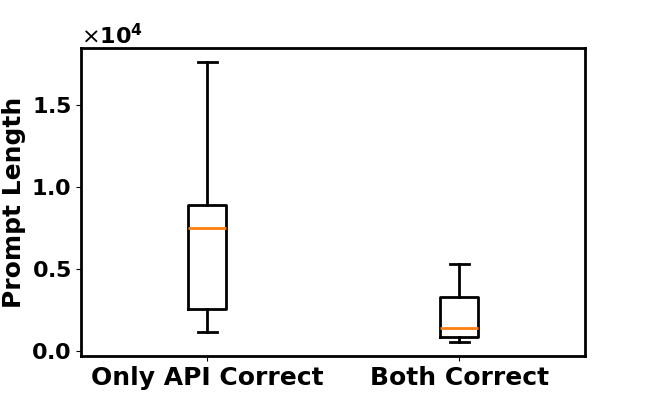}  
    \captionsetup{width=0.95\textwidth}  
    \caption{Comparison of performance based on input prompt length in CodeEval with Gemini}
\end{subfigure}
\caption{Comparison of Input Prompt Lengths for Test Cases: Success in Both ConAPI \& API vs. API-Only Success Across Different Datasets and LLM.}
\label{fig:length}
\end{figure*}

\subsection{RQ2: Can the performance gains from contextual information and invoked APIs be fully captured by simply integrating them?}

To further investigate the effectiveness of combining contextual information with invoked APIs, we analyze the intersection of passed test cases among $\rm Context_{LLM}$, $\rm API_{LLM}$, and $\rm ConAPI_{LLM}$.

To mitigate the impact of this randomness on our experimental results, we evaluate the intersection of passed test cases using the $\rm Pass@5$ metric, ensuring that each test case is tested five times. The experimental results are presented in Figure~\ref{fig:intersection}.

From the figure, we observe that $\rm ConAPI_{LLM}$ successfully covers most of the test cases that passed in either $\rm API_{LLM}$ or $\rm Context_{LLM}$. This result is intuitive, as $\rm ConAPI_{LLM}$ integrates both contextual and API-related information. If the retrieved contextual and API-related information are indeed effective, then $\rm ConAPI_{LLM}$ should naturally pass the same test cases as $\rm Context_{LLM}$ and $\rm API_{LLM}$.

However, we also notice certain test cases that pass in $\rm Context_{LLM}$ or $\rm API_{LLM}$ but fail in $\rm ConAPI_{LLM}$. One possible explanation is the inherent randomness in the model’s output. Despite testing each case five times, some degree of randomness persists, leading to minor inconsistencies in the results. The number of test cases passed by $\rm Context_{LLM}$ but failed by $\rm ConAPI_{LLM}$ is relatively small, suggesting that model randomness might be the primary cause.

In contrast, Figure~\ref{fig:intersection:code_gpt}, Figure~\ref{fig:intersection:repo_gemini}, and Figure~\ref{fig:intersection:code_gemini} reveal a notable number of test cases that pass in $\rm API_{LLM}$ but fail in $\rm ConAPI_{LLM}$. This discrepancy cannot be solely attributed to model randomness, indicating that other factors may be influencing the performance.

To further investigate this issue, we analyze the input prompt length of test cases that were passed by both $\rm API_{LLM}$ and $\rm ConAPI_{LLM}$, as well as those passed only by $\rm API_{LLM}$. The results are shown in Figure~\ref{fig:length}. From this figure, we observe a significant difference in input prompt length between these two sets of test cases. Specifically, the test cases passed only by $\rm API_{LLM}$ tend to have much longer input prompts compared to those passed by both $\rm API_{LLM}$ and $\rm ConAPI_{LLM}$.

Since contextual information is naturally connected to the user query, we structure the prompt by first providing relevant API information, followed by contextual information along with the user query. This ordering results in API-related content appearing farther from the user query in the prompt. While modern LLMs can handle long inputs, content appearing earlier in the prompt may have a diminished impact on later generated responses if the input length is too extensive.

\summaryc{\textbf{Finding 2:} Naively appending invoked APIs with contextual information may cause test cases that could pass with a single piece of information to fail due to excessive input length.}

\begin{table*}[t]
    \setlength\tabcolsep{8pt}
    \centering
    \caption{Statistics of passed test cases: comparison by context alone vs. context with API under different API containment conditions.}
    \label{tab:preiliminary}
    \tabmargin
    \begin{tabular}{lllllllll}
    \toprule    
    \multirow{3}{*}{\textbf{LLM}} & \multicolumn{4}{c}{\textbf{RepoExec}} & \multicolumn{4}{c}{\textbf{CodeEval}}\\
    \cmidrule(l){2-5}
    \cmidrule(l){6-9}
    & \multicolumn{2}{c}{\textbf{Fully Contained}} & \multicolumn{2}{c}{\textbf{Not Included}} & \multicolumn{2}{c}{\textbf{Fully Contained}} & \multicolumn{2}{c}{\textbf{Not Included}}   \\
    \cmidrule(r){2-3} 
    \cmidrule(l){4-5}
    \cmidrule(l){6-7} 
    \cmidrule(l){8-9}
    & \textbf{CPass} & \textbf{BPass} & \textbf{CPass} & \textbf{BPass} & \textbf{CPass} & \textbf{BPass} & \textbf{CPass} & \textbf{BPass} \\
    \midrule
        $\rm GPT$ & 30 (28.0\%) & 18 (60.0\%) & 36 (24.5\%) & 12 (33.3\%) & 8 (47.1\%) & 5 (62.5\%) & 29 (20.3\%) & 8 (27.6\%) \\
        $\rm Gemini$ & 22 (20.6\%) & 15 (68.2\%) & 25 (16.7\%) & 8 (32.0\%) & 4 (22.2\%) & 2 (50.0\%) & 15 (10.3\%) & 2 (13.3\%) \\
    \bottomrule
    \end{tabular}
    \label{tab:coverage}
\end{table*}

\subsection{RQ3: Are the performance improvements from contextual information and invoked APIs independent of each other?}

Since the contextual information contains numerous member functions, some of which may be the invoked functions, it is possible that the contextual information already includes API details. To better understand how different types of information affect model performance in repository-level code generation, we analyze the number of test cases passed by $\rm Context_{LLM}$ under two conditions: (1) when the contextual information fully contains the invoked APIs and (2) when it does not contain any APIs at all. The statistical results are presented in Table~\ref{tab:coverage}.

In this table, $\rm Fully\ Contained$ refers to cases where the contextual information includes all invoked API details, while $\rm Not\ Included$ indicates cases where the contextual information does not contain any invoked APIs. The metric $\rm CPass$ represents the number of test cases passed by $\rm Context_{LLM}$, whereas $\rm BPass$ denotes the test cases passed by both $\rm Context_{LLM}$ and $\rm API_{LLM}$. The percentage under $\rm CPass$ indicates the ratio of passed test cases to the total number of test cases in each condition (Fully Contained or Not Included). The percentage under $\rm BPass$ represents the proportion of test cases passed by $\rm Context_{LLM}$ that are also successfully passed by $\rm API_{LLM}$. To mitigate the effect of model randomness, we attempt to pass each test case five times. To isolate the influence of LLMs' intrinsic code generation capability, we exclude all test cases that can be trivially passed by $\rm Pure_{LLM}$ from our statistical results.

From Table~\ref{tab:coverage}, we observe that API information within the contextual data significantly improves performance. Specifically, by comparing cases where the contextual information fully contains the invoked APIs to those where it does not contain any, we find that $\rm Context_{LLM}$ achieves a notable performance boost in both datasets. In the CoderEval dataset, the performance is even doubled.

Moreover, test cases passed by both $\rm Context_{LLM}$ and $\rm API_{LLM}$ exhibit strong similarity. Notably, when the contextual information fully contains the invoked APIs, approximately 50–70\% of the test cases passed by $\rm Context_{LLM}$ are also passed by $\rm API_{LLM}$. In this scenario, the invoked API information is essentially a subset of the contextual information, suggesting that comparable performance can be achieved by reducing the complexity of contextual information to only the invoked API details. This finding highlights the critical role of invoked APIs in repository-level code generation.

Additionally, we acknowledge that contextual information still provides valuable content beyond the invoked APIs. This is evident from the fact that $\rm Context_{LLM}$ maintains a certain level of test case success even when it does not contain any invoked APIs, while pure API information alone is insufficient for generating correct code. This indicates that contextual information contributes additional useful signals to the code generation process.

\summaryc{\textbf{Finding 3:} Contextual information sometimes includes the invoked APIs for the target code, and a portion of the performance improvement attributed to contextual information actually stems from the presence of these invoked APIs.}

%% file: sections/methodology.tex
\section{\method}
\subsection{Overview}
\begin{figure}
    \centering
    \includegraphics[width=\linewidth]{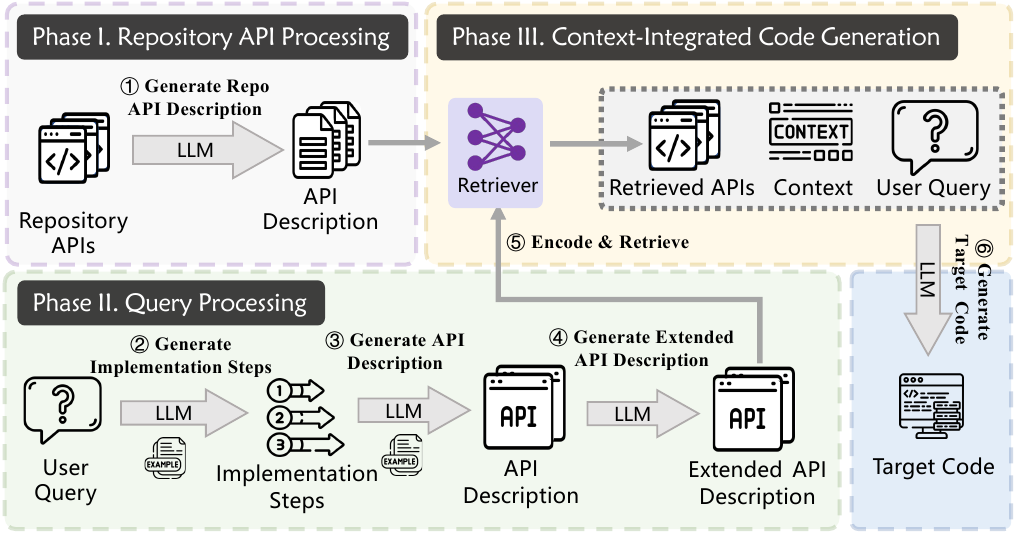}
    \figmargin
    \caption{\method framework.}
    \label{fig:overview}
\end{figure}
Based on our empirical study, potential API retrieval plays a crucial role in repository-level code generation, yet identifying which APIs can be invoked within a repository remains a challenging problem. To address this, we propose \method, a simple yet effective approach for retrieving relevant APIs. As illustrated in Figure~\ref{fig:overview}, \method consists of three phases: repository API processing, query processing, and context-integrated code generation. In the repository API processing phase, we leverage LLMs to generate natural language descriptions for each API in the repository and encode them into representation vectors using pre-trained models. During query processing, we guide LLMs with examples to generate descriptions of potentially invoked API functionalities, which are similarly encoded into vectors. Finally, in the context-integrated code generation phase, we retrieve relevant APIs for each API description based on the cosine similarity between their vector representations. The examples and prompts used in \method are provided in the \href{https://anonymous.4open.science/r/AllianceCoder}{Appendix}.

\subsection{Repository API Processing}
\subsubsection{Generate Repo API Description.}
In our proposed approach, API retrieval is performed using vector cosine similarity, with the encoding process handled by a pre-trained model in an unsupervised manner. However, a semantic gap exists between programming languages and natural language, which can result in low cosine similarity between code snippets and their corresponding descriptions. To mitigate this issue, we first leverage LLMs to generate natural language descriptions for each API in the repository. The process of generating these descriptions can be formulated as follows:
\begin{equation} 
P(D \mid C) = \prod_{t=1}^{T} P(d_t \mid d_{<t}, C)
\end{equation}
\noindent where $D$ represents the code description, $C$ denotes the given code, $T$ is the length of the description, and $P(\cdot)$ represents the probability distribution.

Once the descriptions for all APIs are generated, they are encoded into representation vectors and stored in advance for efficient retrieval.

\subsection{Query Processing}
\subsubsection{Generate Implementation Steps.} The goal of the query processing phase is to generate descriptions of the APIs that may be invoked in the target function. To achieve this, we first provide examples and instruct LLMs to decompose the overall code task into a sequence of concrete implementation steps. This process can be formulated as follows:
\begin{equation} 
\{s_1, \dots, s_n\} = \textit{Generate}(q, \{e_1, \dots, e_m\})
\end{equation}
\noindent where $\{s_1, \dots, s_n\}$ represents the detailed implementation steps generated by LLMs, $q$ is the user query specifying the target code functionality, $\{e_1, \dots, e_m\}$ denotes the examples provided for in-context learning, and $\textit{Generate}(\cdot)$ is the function that integrates the examples and user query to generate the implementation steps.

\subsubsection{Generate API Description.} Once these steps are obtained, they are fed back into the LLM along with additional examples and prompts to generate descriptions of the potential APIs used in each step. This process is expressed as follows:
\begin{equation} 
\{a_1, \dots, a_o\} = \textit{Generate}(\{e_1, \dots, e_m\}, \{s_1, \dots, s_n\})
\end{equation}
\noindent where $\{s_1, \dots, s_n\}$ are the previously generated implementation steps, $\{a_1, \dots, a_o\}$ represents the descriptions of APIs that may be used in these steps, $\{e_1, \dots, e_m\}$ are the examples provided for in-context learning, and $\textit{Generate}(\cdot)$ is the function that integrates the examples and implementation steps to generate API descriptions.

\subsubsection{Generate Extended API Description.} However, LLMs may sometimes generate API descriptions that encompass composite functionalities, requiring multiple APIs for full implementation. To address this issue, we further guide LLMs to expand the set of potential APIs, ensuring broader coverage and reducing the risk of omitting relevant APIs. This expansion process is formulated as follows:
\begin{equation} 
\{a^{\prime}_1, \dots, a^{\prime}_k\} = \textit{Generate}(\{a_1, \dots, a_o\})
\end{equation}
\noindent where $\{a_1, \dots, a_o\}$ are the previously generated potential API descriptions, $\{a^{\prime}_1, \dots, a^{\prime}_k\}$ represents the expanded set of API descriptions, and $\textit{Generate}(\cdot)$ is the function that refines and extends the API descriptions based on the initial candidates.

\subsection{Context-Integrated Code Generation}
\subsubsection{Encode \& Retrieve.} Once the potential API descriptions are generated, they are encoded into representation vectors using the same pre-trained model applied for encoding API descriptions in the repository. API retrieval is then conducted based on vector cosine similarity, ensuring that each API description retrieves only the most relevant API from the repository. 

\subsubsection{Generate Target Code.} Finally, all retrieved APIs are appended to the contextual information and the user’s query before being fed into LLMs.

%% file: sections/evaluation.tex
\section{Evaluation}
We aim to answer the following research questions (RQs):

\begin{itemize}
\item \textbf{RQ4: How effective is \method in repository-level code generation?}

\item \textbf{RQ5: How does \method perform in API prediction, and what is the gap between its predictions and the ideal outcomes?}

\item \textbf{RQ6: How effective is generating natural language descriptions for API retrieval compared to using code snippets?}
\end{itemize}

\begin{table*}[t]
    \setlength\tabcolsep{7pt}
    \centering
    \caption{Performance comparison of different approaches with various LLMs (best performance in bold).}
    \label{tab:overall}
    \tabmargin
    \begin{tabular}{lllllll}
    \toprule
    \multirow{2}{*}{\textbf{Approach}} & \multicolumn{3}{c}{\textbf{RepoExec}} & \multicolumn{3}{c}{\textbf{CoderEval}}\\
    \cmidrule(r){2-4} 
    \cmidrule(l){5-7}
    & \textbf{Pass@1} & \textbf{Pass@3} & \textbf{Pass@5} & \textbf{Pass@1} & \textbf{Pass@3} & \textbf{Pass@5} \\
    \midrule
        $\rm RepoCoder_{GPT}$ & 19.72 & 23.10 & 25.91 & 14.35 & 17.39 & 19.13\\
        $\rm RLCoder_{GPT}$ & 31.55 & 36.62 & 38.87 & 16.52 & 19.13 & 21.30 \\
        $\rm Context_{GPT}$ & 29.01 & 34.08 & 38.59 & 30.43 & 36.52 & 37.83 \\
        $\rm \method_{GPT}$ & \textbf{34.93 (10.7\%$\uparrow$)} & \textbf{40.28 (10.0\%$\uparrow$)} & \textbf{41.97 (8.0\%$\uparrow$)} & \textbf{36.52 (20.0\%$\uparrow$)} & \textbf{40.00 (9.5\%$\uparrow$)} & \textbf{41.30 (9.2\%$\uparrow$)} \\
        \hdashline
        $\rm ConAPI_{GPT}$ & 37.75 & 44.51 & 47.04 & 36.52 & 41.30 & 41.74 \\
    \midrule
        $\rm RepoCoder_{Gemini}$ & 25.63 & 31.26 & 32.95 & 13.91 & 15.65 & 16.09\\
        $\rm RLCoder_{Gemini}$ & 12.68 & 15.21 & 18.31 & 12.17  & 12.61 & 13.04 \\
        $\rm Context_{Gemini}$ & 30.99 & 34.08 & 35.77 &23.04 & 26.09 & 27.39 \\
        $\rm \method_{Gemini}$ & \textbf{35.49 (14.5\%$\uparrow$)}  & \textbf{39.16 (14.9\%$\uparrow$)}  & \textbf{41.13 (15.0\%$\uparrow$)}  & \textbf{24.78 (7.6\%$\uparrow$)} & \textbf{26.52 (1.6\%$\uparrow$)} & \textbf{27.82 (1.6\%$\uparrow$)}\\
        \hdashline
        $\rm ConAPI_{Gemini}$ & 38.59 & 41.69 & 44.51 & 25.22 & 27.39 & 28.70 \\
    \bottomrule
    \end{tabular}
    \label{tab:overall}
\end{table*}

\begin{table*}[t]
    \setlength\tabcolsep{7pt}
    \centering
    \caption{Performance comparison between text-to-text and text-to-code retrieval.} \label{tab:text2code}
    \tabmargin
    \begin{tabular}{lllllll}
    \toprule
    \multirow{2}{*}{\textbf{Approach}} & \multicolumn{3}{c}{\textbf{RepoExec}} & \multicolumn{3}{c}{\textbf{CoderEval}}\\
    \cmidrule(r){2-4} 
    \cmidrule(l){5-7}
    & \textbf{Pass@1} & \textbf{Pass@3} & \textbf{Pass@5} & \textbf{Pass@1} & \textbf{Pass@3} & \textbf{Pass@5} \\
    \midrule
    $\rm \method_{GPT}$ & 34.93 & 40.28 & 41.97 & 36.52 & 40.00 & 41.30 \\
    $\rm \method (Code)_{GPT}$ & 33.71 (3.5\%$\downarrow$) & 39.38 (2.2\%$\downarrow$) & 41.93 (0.1\%$\downarrow$) & 23.47 (35.7\%$\downarrow$) & 25.65 (35.9\%$\downarrow$) & 26.96 (34.7\%$\downarrow$) \\
    \midrule
    $\rm \method_{Gemini}$ & 35.49 & 39.16 & 41.13  & 24.78 & 26.52 & 27.82\\
    $\rm \method (Code)_{Gemini}$ & 28.45 (19.8\%$\downarrow$) & 33.80 (13.7\%$\downarrow$) & 34.65 (15.8\%$\downarrow$) & 18.26 (26.3\%$\downarrow$) & 20.43 (23.0\%$\downarrow$) & 21.30 (23.4\%$\downarrow$) \\
    \bottomrule
    \end{tabular}
    \label{tab:code}
\end{table*}

\subsection{RQ4: How effective is \method in repository-level code generation?}

Table~\ref{tab:overall} presents a performance comparison of different approaches across various LLMs. As a baseline, we include $\rm Context_{LLM}$ from our preliminary study. The results demonstrate that \method consistently achieves state-of-the-art (SOTA) performance across all evaluation metrics on both datasets.

Notably, \method exhibits a substantial improvement in the $\rm Pass@1$ metric, with an increase of approximately 10-20\%. Since $\rm Pass@1$ reflects the success rate on the first attempt, this enhancement significantly reduces the need for multiple regeneration attempts, thereby improving the overall user experience.

Furthermore, the performance gains of \method are particularly pronounced compared to $\rm Context_{LLM}$ and approach those of $\rm ConAPI_{GPT}$ on the CoderEval dataset. This result underscores the effectiveness of API retrieval in \method. However, it is important to note that the performance improvement of \method on the CoderEval dataset with Gemini is relatively limited. This is primarily due to the inherently low upper bound on the potential gains from API retrieval. Even under optimal API retrieval conditions, the maximum achievable improvement in $\rm Pass@3$ and $\rm Pass@5$ remains around 5\%. We attribute this limitation to the intrinsic generation capabilities of the LLM, which constrain the extent to which API retrieval can enhance performance.

\summaryb{\textbf{Summary 1:} \method achieves SOTA performance across all baselines, with particularly notable improvements in $\rm Pass@1$. On the CoderEval dataset with GPT, \method's performance closely approaches the theoretical upper bound of API retrieval.}

\subsection{{RQ5: How does \method perform in API prediction, and what is the gap between its predictions and the ideal outcomes?}}

\begin{table}[t]
    \setlength\tabcolsep{2pt}
    \centering
    \small
    \caption{Comparison of API recall counts: \method vs. actual API counts.} \label{tab:text2code}
    \tabmargin
    \begin{tabular}{lllllll}
    \toprule
    \multirow{2}{*}{\textbf{Approach}} & \multicolumn{3}{c}{\textbf{RepoExec}} & \multicolumn{3}{c}{\textbf{CoderEval}}\\
    \cmidrule(r){2-4} 
    \cmidrule(l){5-7}
    & \textbf{Higher} & \textbf{Equal} & \textbf{Lower} & \textbf{Higher} & \textbf{Equal} & \textbf{Lower} \\
    \midrule
    $\rm \method_{GPT}$ & 74.45 & 6.9 & 18.61 & 72.64 & 4.7 & 22.64 \\
    $\rm \method_{Gemini}$ & 51.42 & 13.25 & 35.33 & 58.49 & 9.9 & 31.60 \\
    \bottomrule
    \end{tabular}
    \label{tab:count}
\end{table}

\begin{table}[t]
    \setlength\tabcolsep{1.5pt}
    \centering
    \small
    \caption{Comparison of API recall ratios for \method and \method with context combination.} \label{tab:text2code}
    \tabmargin
    \begin{tabular}{lllllll}
    \toprule
    \multirow{2}{*}{\textbf{Approach}} & \multicolumn{3}{c}{\textbf{RepoExec}} & \multicolumn{3}{c}{\textbf{CoderEval}}\\
    \cmidrule(r){2-4} 
    \cmidrule(l){5-7}
    & \textbf{Recall} & \textbf{BRecall} & \textbf{CRecall} & \textbf{Recall} & \textbf{BRecall} & \textbf{CRecall} \\
    \midrule
    $\rm \method_{GPT}$ & 20.38 & 21.33 & 29.23 & 16.81 & 24.00 & 46.22 \\
    $\rm \method_{Gemini}$ & 14.62 & 16.05 &23.46 & 10.08 & 15.38 & 39.50 \\
    \bottomrule
    \end{tabular}
    \label{tab:recall}
\end{table}

To evaluate the effectiveness of \method in API prediction, we first examine the API recall numbers obtained by \method, as shown in Table~\ref{tab:count}. In this table, $\rm Higher$ represents the percentage of test cases where \method recalls more APIs than the actual number required by the target function, $\rm Equal$ represents the percentage of test cases where \method recalls exactly the same number of APIs as the actual requirement, and $\rm Lower$ represents the percentage of test cases where \method recalls fewer APIs than needed. From the table, we observe that LLMs tend to generate more APIs than the target function actually requires. Additionally, different LLMs exhibit varying tendencies; for example, GPT tends to recall more APIs compared to Gemini.

Table~\ref{tab:recall} presents the API recall ratios for \method and \method with context combination. In this table, $\rm Recall$ represents the API recall ratio of \method across all test cases, $\rm BRecall$ represents the API recall ratio for test cases that are successfully passed by both \method and $\rm ConAPI_{LLM}$ (i.e., the ideal condition with API retrieval), and $\rm CRecall$ represents the API recall ratio of \method, including APIs invoked in the contextual information. From the results, we observe that the API recall ratio for passed test cases is higher than the overall recall ratio of \method, demonstrating that providing correct potential APIs as input can enhance performance. However, even when considering APIs embedded in contextual information, the API recall ratio remains relatively low. Nevertheless, the overall performance of \method is close to the ideal performance achieved by $\rm ConAPI_{LLMs}$, indicating that not only does providing all invoked APIs improve model performance, but even offering a subset of invoked APIs can contribute to performance enhancement.

\summaryb{\textbf{Summary 2:} \method tends to predict more APIs than actually required, and providing LLMs with a subset of potential invoked APIs can also enhance repository-level code generation performance.}

\subsection{RQ6: How effective is generating natural language descriptions for API retrieval compared to using code snippets?}

Although most existing pre-trained models attempt to unify programming languages and natural language modalities during the pre-training stage in an unsupervised manner, a persistent alignment gap remains between these two modalities. This misalignment results in representation vectors of code-query pairs that do not align well, which serves as the primary motivation for \method’s approach—first generating natural language descriptions for APIs before retrieval.

To examine the impact of this semantic gap on overall performance, we conduct an experiment in which APIs from the repository are directly encoded into representation vectors, followed by retrieval based on these vectors. The results, summarized in Table~\ref{tab:code}, yield several key observations.

First, performance is highly sensitive to the specific implementation details of code within the repository. Notably, the performance degradation in the CoderEval dataset is significantly more pronounced than in the RepoExec dataset for both LLMs. We attribute this to the greater functional complexity and diverse coding styles in CoderEval, which exacerbate the semantic gap between code and its corresponding natural language descriptions.

Additionally, the ability of LLMs to generate accurate and semantically meaningful descriptions plays a crucial role in retrieval performance. Different LLMs exhibit varying degrees of performance degradation within the same dataset. For instance, in RepoExec, GPT suffers a smaller performance drop compared to Gemini, whereas in CoderEval, GPT performs worse. This discrepancy likely arises from differences in description generation styles. Since each LLM employs distinct training strategies and datasets, their output formats vary, ultimately influencing retrieval effectiveness.

\summaryb{\textbf{Summary 3:} Directly using code for retrieval leads to a decline in \method’s overall performance. The extent of this performance drop varies depending on the choice of LLM and dataset.}

%% file: sections/related_works.tex
\section{Related Works}
\subsection{Code Generation}
The task of repository-level code
generation is gaining significant attention for intelligent software
development in real-world scenarios~\cite{liu2021opportunities, svyatkovskiy2020intellicode, wang2021code, zan2022cert}. Traditional code generation methods can be primarily categorized into rule-based approaches~\cite{hindle2016naturalness}, statistical probability model-based methods~\cite{raychev2014code}, and deep learning-based approaches~\cite{bhoopchand2016learning, izadi2022codefill, wang2021code, chen2023api}. With the advancement of large language models~\cite{achiam2023gpt, liu2024deepseek, yang2024qwen2, guo2024deepseek, roziere2023code}, many researchers have introduced LLMs into code generation tasks~\cite{guo2024stop, zhu2024hot, li2024ircoco, yoran2023making}. In combination with LLMs, numerous studies have adopted Retrieval-Augmented Generation techniques for code completion and code generation tasks~\cite{lu2022reacc, hayati2018retrieval, parvez2021retrieval, zhou2022docprompting, zan2023private}. 

For examples, ReCode~\cite{hayati2018retrieval} improves Neural Code Generation through subtree retrieval. RedCoder~\cite{parvez2021retrieval} proposes to first retrieve top-k candidate codes for a given code functionality description, then aggregate these candidates, and finally generate the target code. DocPrompting~\cite{zhou2022docprompting} adopts code documentation to improve code generation, addressing the challenges of generating unknown functions and library code. ReACC~\cite{lu2022reacc} retrieves similar code from a code database, scores retrieved code snippets using weighted results of cosine similarity and the BM25, and concatenates the final retrieved results with incomplete code as input to the LLM. This series of studies demonstrates that combining LLM and RAG techniques significantly enhances the performance of code completion and code generation. APICoder~\cite{zan2023private}trains models through API documentation to better generate private library code. 

\subsection{Repository-level Code Generation}
Repository-level code generation, which leverages the extensive context available across an code repository, has emerged as a central research focus in the field. Studies have attempted to improve repository-level code generation~\cite{zhang2023repocoder, eghbali2024hallucinator, ding2022cocomic, phan2024repohyper, liu2024graphcoder, wang2024rlcoder}. 
RepoCoder~\cite{zhang2023repocoder} and De-Hallucinator~\cite{eghbali2024hallucinator} adopt an iterative retrieval and generation framework.
RepoMinCoder~\cite{li2024repomincoder} builds upon the traditional retrieval-augmented generation method by introducing an additional round of screening and ranking based on information loss.
R2C2-Coder~\cite{deng2024r2c2} first constructs a candidate retrieval pool and then retrieves relevant content from the pool for each completion position, assembling it into a completion prompt.
The aforementioned methods extract the target code for retrieval. However, they overlook the intrinsic structure within the code. To address this, CoCoMIC~\cite{ding2022cocomic} and RepoHyper~\cite{phan2024repohyper}construct method-level graphs to enhance the retrieval process. 
Nevertheless, these methods fail to capture statement-level structures, which are crucial for understanding code semantics. GraphCoder~\cite{liu2024graphcoder} overcomes this limitation by incorporating statement-level structural information. RepoGraph~\cite{ouyang2024repograph} is a framework that converts repositories into graph structures to enhance code analysis and understanding. Rambo~\cite{bui2024rambo} identifies and incorporates repository-specific elements and their usages to generate more accurate method bodies in large codebases. DraCo~\cite{cheng2024dataflow} uses extended dataflow analysis to create a repo-specific context graph for more precise retrieval of relevant background knowledge.
CodePlan~\cite{bairi2024codeplan}, RepoFuse~\cite{liang2024repofuse}, and 
A\textsuperscript{3}CodeGen~\cite{liao20243} leverage static code analysis to identify and retrieve relevant candidate code.
However, methods that use a fixed window risk losing code semantics, while dependency parsing approaches are restricted to a limited context within the dependency graph, making them ineffective in complex scenarios. To overcome these challenges, the researchers introduces RLCoder~\cite{wang2024rlcoder}, a repository-level code genration method driven by reinforcement learning. RepoGenReflex~\cite{wang2024repogenreflex} is another framework driven by reinforcement learning. CoCoGen~\cite{bi2024iterative} uses compiler feedback and static analysis to identify and fix mismatches between generated code and project-specific context. RepoGenix~\cite{RepoGenix} combines analogous and relevant contexts with Context-Aware Selection to compress information into compact prompts.

In methods such as KNN-LM~\cite{khandelwal2019generalization}, KNM-LM~\cite{tang2023domain}, and FT2Ra~\cite{guo2024ft2ra}, a retrieval process is triggered for each generated token. Consequently, as the generated sequence length increases,  leading to a significant increase in retrieval time.
CodeAgent~\cite{zhang2024codeagent} and ToolGen~\cite{wang2024toolgen} investigate the integration of tool invocation mechanisms.
Additionally, benchmarks such as RepoEval~\cite{zhang2023repocoder}, RepoBench~\cite{liu2023repobench}, and CrossCodeEval~\cite{ding2023crosscodeeval} have been introduced to systematically evaluate various code generation capabilities across different contexts, thereby driving progress in this field. Some work study selective retrieval~\cite{wu2024repoformer,zhang2024lightweight}. RepoFormer~\cite{wu2024repoformer} improves repository-level code generation by deciding when retrieval is necessary. CARD~\cite{zhang2024lightweight} is a lightweight critique method that optimizes retrieval-augmented mechanism by determining when retrieval is necessary and selecting the best prediction, reducing retrieval frequency. Probing-RAG~\cite{baek2024probing} is a system that analyzes models' hidden states to adaptively determine when to retrieve external knowledge.

In summary, these approach proves to be effective for repository-level code generation. Nevertheless, identifying which APIs can be invoked within a repository remains a challenging problem. To tackle this issue, we first generate descriptions for each API in the repository and encode them into vectors. Furthermore, we employ in-context learning and chain-of-thought prompting to assist LLMs in generating descriptions of potentially relevant API functionalities, which are then encoded into vectors.  By computing cosine similarity between these vectors, we identify the most relevant APIs, enabling efficient and precise retrieval that improves code generation outcomes.

\section{Threats to Validity}
\paragraph{\textbf{Internal Threats}}  
The first internal threat relates to the scope of our experimental datasets. While we evaluated our approach on Python programming language. This may limit the generalizability of our findings to other programming languages. To mitigate this, we used CoderEval and RepoExec benchmarks which emphasize real-world repositories. In the future, we will conduct experiments on more progarmming languages.

The second internal threat stems from the selection of LLMs. Due to computational constraints, we evaluated only two LLMs (GPT and Gemini variants), which may not fully represent the capabilities of all state-of-the-art models. However, these models were chosen for their dominance in code generation research~\cite{ouyang2024repograph, zhang2024codeagent,bui2024rambo}, and our methodology is model-agnostic by design.  

The third internal threat involves our evaluation metric. While Pass@k effectively measures functional correctness, it does not assess code quality aspects such as readability, maintainability, or alignment with repository conventions. Future work could incorporate static analysis or human evaluations to address this threat.  

\paragraph{\textbf{External Threats}}  
The first external threat is potential data leakage. Although our datasets are widely used in prior work, some repository code may have been included in the pretraining data of evaluated LLMs. 
In the future, dynamically evolving benchmarks could be investigated to mitigate this threat.

The second external threat is the inherent randomness in LLM generation. Despite setting temperature to 0.7, minor output variations may persist. We mitigated this by  repeating experiments five times per test case and reporting averaged \textit{Pass@k} scores.  

The third external threat concerns the generalizability of our retrieval strategy. \method relies on unsupervised API description generation, which assumes APIs have meaningful natural language semantics. This may underperform in repositories with poorly documented or cryptically named APIs. We partially addressed this by using LLM-generated descriptions to bridge semantic gaps, but edge cases (e.g., generated hallucinated descriptions) require further investigation.  

\section{Conclusion}
This paper presents an empirical study on the role of retrieved information in retrieval-augmented repository-level code generation. Empirical results demonstrate that in-context code and potential API information significantly enhance LLM performance, while retrieved similar code snippets often introduce noise. Based on these insights, we propose \method, a novel context-integrated method that leverages chain-of-thought prompting to decompose requirements and retrieve APIs via semantic descriptions according to the decomposed requirements. Experiments on multiple benchmarks show that \method outperforms state-of-the-art baselines by up to 20\% in Pass@1, highlighting the importance of targeted retrieval strategies for complex code generation. 